\documentstyle[12pt]{article}
\catcode`\@=11

\def\@normalsize{\@setsize\normalsize{15pt}\xiipt\@xiipt
\abovedisplayskip 14pt plus3pt minus3pt%
\belowdisplayskip \abovedisplayskip
\abovedisplayshortskip  \z@ plus3pt%
\belowdisplayshortskip  7pt plus3.5pt minus0pt}
\def\small{\@setsize\small{13.6pt}\xipt\@xipt
\abovedisplayskip 13pt plus3pt minus3pt%
\belowdisplayskip \abovedisplayskip
\abovedisplayshortskip  \z@ plus3pt%
\belowdisplayshortskip  7pt plus3.5pt minus0pt
\def\@listi{\parsep 4.5pt plus 2pt minus 1pt
            \itemsep \parsep
            \topsep 9pt plus 3pt minus 3pt}}

\def\underline#1{\relax\ifmmode\@@underline#1\else
        $\@@underline{\hbox{#1}}$\relax\fi}
\@twosidetrue
\relax

\catcode`@=12

\catcode`\@=11

\def\section{\@startsection{section}{1}{\z@}{3.5ex plus 1ex minus
   .2ex}{2.3ex plus .2ex}{\large\bf}}

\def\ps@headings{\def\@oddfoot{}\def\@evenfoot{}
\def\@oddhead{\hbox{}\hfill
        \makebox[.5\textwidth]{\raggedright\ignorespaces --\thepage{}--
        \hfill }}
\def\@evenhead{\@oddhead}
\def\subsectionmark##1{\markboth{##1}{}}
}

\ps@headings

\catcode`\@=12

\relax

\def\figcap{\section*{Figure Captions\markboth
        {FIGURECAPTIONS}{FIGURECAPTIONS}}\list
        {Fig. \arabic{enumi}:\hfill}{\settowidth\labelwidth{Fig. 999:}
        \leftmargin\labelwidth
        \advance\leftmargin\labelsep\usecounter{enumi}}}
 \relax
\def\tablecap{\section*{Table Captions\markboth
        {TABLECAPTIONS}{TABLECAPTIONS}}\list
        {Table \arabic{enumi}:\hfill}{\settowidth\labelwidth{Table 999:}
        \leftmargin\labelwidth
        \advance\leftmargin\labelsep\usecounter{enumi}}}
 \relax
\def\reflist{\section*{References\markboth
        {REFLIST}{REFLIST}}\list
        {[\arabic{enumi}]\hfill}{\settowidth\labelwidth{[999]}
        \leftmargin\labelwidth
        \advance\leftmargin\labelsep\usecounter{enumi}}}
 \relax

\catcode`\@=11

\def\marginnote#1{}
\newcount\hour
\newcount\minute
\newtoks\amorpm
\hour=\time\divide\hour by60
\minute=\time{\multiply\hour by60 \global\advance\minute by-
\hour}
\edef\standardtime{{\ifnum\hour<12 \global\amorpm={am}%
    \else\global\amorpm={pm}\advance\hour by-12 \fi
    \ifnum\hour=0 \hour=12 \fi
    \number\hour:\ifnum\minute<100\fi\number\minute\the\amorpm}}
\edef\militarytime{\number\hour:\ifnum\minute<100\fi\number\minute}
\def\draftlabel#1{{\@bsphack\if@filesw {\let\thepage\relax
  \xdef\@gtempa{\write\@auxout{\string
    \newlabel{#1}{{\@currentlabel}{\thepage}}}}}\@gtempa
    \if@nobreak \ifvmode\nobreak\fi\fi\fi\@esphack}
     \gdef\@eqnlabel{#1}}
\def\@eqnlabel{}
\def\@vacuum{}
\def\draftmarginnote#1{\marginpar{\raggedright\scriptsize\tt#1}}
\def\draft{\oddsidemargin -.5truein
        \def\@oddfoot{\sl preliminary draft \hfil
        \rm\thepage\hfil\sl\today\quad\militarytime}
        \let\@evenfoot\@oddfoot \overfullrule 3pt
        \let\label=\draftlabel
        \let\marginnote=\draftmarginnote
   
\def\@eqnnum{(\theequation)\rlap{\kern\marginparsep\tt\@eqnlabel}%
\global\let\@eqnlabel\@vacuum}  }
\def\preprint{\twocolumn\sloppy\flushbottom\parindent 1em
        \leftmargini 2em\leftmarginv .5em\leftmarginvi .5em
        \oddsidemargin -.5in    \evensidemargin -.5in
        \columnsep 15mm \footheight 0pt
        \textwidth 250mmin      \topmargin  -.4in
        \headheight 12pt \topskip .4in
        \textheight 175mm
        \footskip 0pt
        
\def\@oddhead{\thepage\hfil\addtocounter{page}{1}\thepage}
        \let\@evenhead\@oddhead \def\@oddfoot{} \def\@evenfoot{} 
}
\def\titlepage{\@restonecolfalse\if@twocolumn\@restonecoltrue\onecolumn
     \else \newpage \fi \thispagestyle{empty}\c@page\z@
        \def\thefootnote{\fnsymbol{footnote}} }
\def\endtitlepage{\if@restonecol\twocolumn \else  \fi
        \def\thefootnote{\arabic{footnote}}
        \setcounter{footnote}{0}}  
\catcode`@=12
\relax

\def\ps@headings{\def\@oddfoot{}\def\@evenfoot{}
\def\@oddhead{\hbox{}\hfill
        \makebox[.5\textwidth]{\raggedright\ignorespaces --\thepage{}--
        \hfill }}
\def\@evenhead{\@oddhead}
\def\subsectionmark##1{\markboth{##1}{}}
}

\ps@headings

\relax

\def\firstpage#1#2#3#4#5#6{
\begin{document}
\begin{titlepage}
\nopagebreak
\title{\begin{flushright}
        \vspace*{-1.8in}
        {\normalsize CERN-TH/98-320}\\[-9mm]
   {\normalsize DEMO-HEP/98-04}\\[-9mm]
        {\normalsize hep-th/9810042}\\[4mm]
\end{flushright}
\vspace{2cm}
{#3}}
\author{\large #4 \\[0.0cm] #5}
\maketitle
\vskip 5mm
\nopagebreak 
\begin{abstract}
{\noindent #6}
\end{abstract}
\vfill
\begin{flushleft}
\rule{16.1cm}{0.2mm}\\[-3mm]
$^\ast$e-mails: bakas@nxth04.cern.ch, 
bakas@ajax.physics.upatras.gr\\[-3mm]
\hspace{1.45cm}{manolis@timaios.nrcps.ariadne-t.gr} \\[-3mm]
\hspace{1.45cm}{alexandros.kehagias@cern.ch} \\[0mm]
CERN-TH/98-320\\[-1mm]
October  1998
\end{flushleft}
\thispagestyle{empty}
\end{titlepage}}

\def\simlt{\stackrel{<}{{}_\sim}}
\def\simgt{\stackrel{>}{{}_\sim}}
\newcommand{\dal}{\raisebox{0.085cm}
{\fbox{\rule{0cm}{0.07cm}\,}}}

\newcommand{\be}{\begin{eqnarray}}
\newcommand{\ee}{\end{eqnarray}}
\newcommand{\btau}{\bar{\tau}}
\newcommand{\p}{\partial}
\newcommand{\bp}{\bar{\partial}}
\renewcommand{\a}{\alpha}
\renewcommand{\b}{\beta}
\newcommand{\g}{\gamma}
\renewcommand{\d}{\delta}
\newcommand{\gsi}{\,\raisebox{-0.13cm}{$\stackrel{\textstyle
>}{\textstyle\sim}$}\,}
\newcommand{\lsi}{\,\raisebox{-0.13cm}{$\stackrel{\textstyle
<}{\textstyle\sim}$}\,}
\date{}
\firstpage{3118}{IC/95/34}
{\large 
{\Large O}CTONIONIC   {\Large G}RAVITATIONAL
{\Large I}NSTANTONS\\
\phantom{X}}
{I. Bakas$^a$, E.G. Floratos$^{b,c}$
 and  A. Kehagias${^{d}}^\ast$
\phantom{X} }
{\vspace{-.4cm}
\normalsize\sl $^a$
Department of Physics, University of Patras, GR-26500, Greece\\
\vspace{-.4cm}
\normalsize\sl $^b$
Institute of Nuclear Physics, NRCS Democritos, Athens, Greece\\
\vspace{-.4cm}
\normalsize\sl 
$^c$Physics Department, University of Crete, Iraklion, Greece \\
\normalsize\sl $^d$ Theory Division, CERN, 1211 Geneva 23, Switzerland
}
{We construct  eight-dimensional gravitational instantons
by solving  appropriate self-duality equations  for the
spin-connection. The particular gravitational instanton 
we present has $Spin(7)$  holonomy
and, in a sense, it is the eight-dimensional
analog of the Eguchi-Hanson 4D space. 
It has a removable bolt singularity which is topologically $S^4$ and
its boundary at infinity is the squashed $S^7$. 
We also lift our solutions to 
ten and eleven dimensions and construct fundamental string and membrane
configurations that preserve 1/16 of the original supersymmetries.}
\newpage
\section{Introduction}

The main purpose of the present work is to construct an
eight-dimensional analogue of the non-compact gravitational
instantons that arise in four dimensions \cite{H,GHP,EGH}, the simplest
example being the Eguchi-Hanson space \cite{EH}. It is well known 
that in four dimensions the non-trivial structure of 
such gravitational instantons is encoded into the
asymptotic form of the metric whose boundary structure
at infinity is given by the lens spaces of $S^3$. Since $S^3$ is
the Hopf fibration of $S^1$ over $S^2$ it is quite natural
to employ the topological structure of the seven-sphere,
$S^7$, as Hopf fibration of $S^3 = SU(2)$ over 
$S^4$ in our search of eight-dimensional gravitational
instantons. In the context of gauge fields the first
Hopf fibration arises in the construction of the $U(1)$
Dirac monopole with unit magnetic charge, while the
second Hopf fibration arises in the construction of the
$SU(2)$ Yang-Mills instanton with unit topological 
number \cite{T}. Because of this analogy we expect that self-dual 
$SU(2)$ Yang-Mills configurations in four dimensions will 
play an interesting role in future investigations of
eight-dimensional gravitational instantons, for instance
in generalizing the notion of nut potential in eight
dimensions and classifying the fixed point sets of their
isometry groups. We will make a few remarks in this 
direction at the very end of the paper.

In any event the solution we present here could be viewed
as the simplest non-trivial example of a more general
class of eight-dimensional gravitational instantons, but
further results will be reported elsewhere. The metric
we construct in the sequel has the squashed seven-sphere 
(as opposed to the round seven-sphere) as its   
boundary space at infinity, which is 
essential for establishing the non-triviality of our solution
in eight dimensions. Alternatively we could have arrived 
at the same result by deforming the squashed seven-sphere
in the interior of an eight dimensional space using  
appropriately chosen coefficient functions, but we decided  
to emphasize the aspects of Hopf fibration in our  
presentation in view of other future generalizations.

As we have shown in \cite{FK}, eight-dimensional
manifolds which satisfy an appropriate self-duality
condition for the spin connection have  holonomy in $Spin(7)$. This provides
us with a systematic way of constructing $Spin(7)$ holonomy manifolds  by
solving appropriate first-order equations.  
Non-compact such manifolds  have first been constructed in 
\cite{B} and further discussed in \cite{AL}, while 
compact ones were constructed as $T^8$ orbifolds in \cite{J7}
after appropriate resolution of the singularities. Here, 
by solving the eight-dimensional   self-duality 
condition we find a space with  $Spin(7)$ holonomy and  isometry group  
$SO(5)\!\times\! SU(2)$. 
It has a bolt singularity which is topologically $S^4$ 
and its boundary at infinity is the squashed $S^7$. Finally, we lift our 
solution in ten and eleven dimensions and we describe fundamental string 
and membrane configurations, which we name octonionic, 
and which preserve 1/16 
of the original supersymmetries.

\section{Octonionic Algebra}

We will recall here some properties of the octonionic
(Cayley) algebra ${\cal{O}}$ and its relation to $SO(8)$
 following mainly \cite{GG,WN}.
The octonionic algebra is a division 
algebra, which means it has a non-degenerate quadratic form $Q$ that satisfies
$Q(xy)=Q(x)Q(y)$ and in addition $Q(x)=0$  implies $x=0$. The other division
algebras are the real ${\cal R}$, complex ${\cal C}$ and quaternionic 
${\cal H}$ algebras. ${\cal R}$, ${\cal C}$ are commutative, 
${\cal H}$ is non-commutative but associative,  
while ${\cal {O}}$ is neither commutative nor associative.   
A basis for ${\cal{O}}$ is provided by the eight elements
\be
1,~e_a, ~~~~~ a=1,\ldots,7, 
\ee
which satisfy the relation 
\be
e_ae_b=\psi_{abc}e_c-\delta_{ab}\, . \label{Pauli}
\ee
The tensor $\psi_{abc}$ is totally antisymmetric with
\be
\psi_{abc}=+1~~~{\mbox{for}}~~~~abc=123,~516,~624,~435,~471~,673,~572.
\ee
We may also define its dual $\psi_{abcd}$ as
\be
\psi_{abcd}=\frac{1}{3!} \epsilon_{abcdfgh}\psi^{fgh}\, , \label{psi}
\ee
so that 
\be
\psi_{abcd}=+1~~~ {\mbox{for}}~~~~abcd=1245,~2671,~3526,~4273,~5764,~6431,~
7531.
\ee 
They satisfy the relations 
\be
\psi^{abc}\psi_{dhc}&=&\d^a_d\d^b_h-\d^a_h\d^b_d-{\psi^{ab}}_{dh}\, , 
\nonumber\\
\psi^{abcd}\psi_{ehcd}&=&4\left(\d^a_e\d^b_h-\d^a_h\d^b_e\right)-
2{\psi^{ab}}_{eh}\, , \nonumber \\
\psi^{abc}\psi_{debc}&=&-4{\psi^{a}}_{de}\, . 
\ee

Let us note here that the relation (\ref{Pauli}) is similar to the
one obeyed by the quaternions
(Pauli matrices). However, the latter satisfy the Jacobi identity as a 
result of the associativity of the quaternionic algebra, 
while the octonions $e_a$ are not associative
and hence do not satisfy the Jacobi identity.   

The tensor $\psi_{abc}$ can be assigned to an $SO(8)$ representation 
$\Psi_{\a\b\g\d}$, where the Greek indices range from $1$ to $8$ and
also set the notation $\a = (a, 8)$. We have   
\begin{eqnarray}
\Psi_{\a\b\g 8}=\psi_{abc}\, , ~~~~
\Psi_{\a\b\g\d}=\psi_{abcd}\, , \label{p}
\end{eqnarray}
which is self-dual 
\be
\Psi_{\a\b\g\d}=\frac{1}{4!}\epsilon_{\a\b\g\d\zeta\eta\theta\kappa}
\Psi^{\zeta\eta\theta\kappa}\, , \label{pe}
\ee
and belongs to one of the three different ${\bf 35}$'s of $SO(8)$ ${\bf 35}_v$,
${\bf 35}_\pm$ 
(related by triality). 
It satisfies the fundamental identity
\be
\Psi_{\a\b\g\d}\Psi^{\zeta\eta\theta\d}&=&
\big{(}\d_\a^\zeta\d_\b^\eta-\d_\b^\zeta\d_\a^\eta)\d_\g^\theta+
(\d_\a^\theta\d_\b^\zeta-\d_\b^\zeta\d_\a^\theta)\d_\g^\eta+
(\d_\a^\eta\d_\b^\theta-\d_\b^\theta\d_\a^\eta)\d_\g^\eta+
\nonumber \\
&&{\Psi_{\a\b}}^{\zeta\eta}\delta_\g^\theta
+{\Psi_{\a\b}}^{\theta\zeta}\delta_\g^\eta+
{\Psi_{\a\b}}^{\eta\theta}\delta_\g^\zeta +
{\Psi_{\g\a}}^{\zeta\eta}\delta_\b^\theta+
{\Psi_{\g\a}}^{\theta\zeta}\delta_\b^\eta+
\nonumber \\&&{\Psi_{\g\a}}^{\eta\theta}\delta_\b^\zeta+
{\Psi_{\b\g}}^{\zeta\eta}\delta_\a^\theta+
{\Psi_{\b\g}}^{\theta\zeta}\delta_\a^\eta+
{\Psi_{\b\g}}^{\eta\theta}\delta_\a^\zeta \, . \label{fund}
\ee 

We may use the octonions $e_a$  to construct a 
 representation of the $SO(7)$ $\g$-matrices according to 
\be
\left(\gamma_a\right)_{bc}=i\psi_{abc}\, , ~~~\left(\gamma_a\right)_{b8}=i
\delta_{ab}\, , ~~~~~\{\gamma_a,\gamma_b\}=2\delta_{ab}\, ,
\ee
so that $\g^{ab}=\frac{1}{2}[\g^a,\g^b]$ are the $SO(7)$ generators. 
We may also form the $SO(8)$ $\g$-matrices $\Gamma_\a=(\Gamma_a,\Gamma_8)\,, 
\a=1,\ldots,8$
\be
\Gamma_a=\left(\matrix{0&i\gamma_a\cr -i\gamma_a&0}\right)\, , 
~~~~~~ \Gamma_8=
\left(\matrix{0&1\cr 1&0}\right)\, , ~~~~~~~
\{\Gamma_\a,\Gamma_\b\}=2\d_{\a\b}\, , 
\ee
that correspond to the standard embedding 
of $SO(7)_v$ in $SO(8)$. 
The latter is defined as the stability subgroup  $SO(7)\subset 
SO(8)$ of the vector 
representation according to which we have the decomposition
\be
{\bf 8}_v={\bf 7}+{\bf 1}\, , ~~~{\bf 8}_{\pm}={\bf 8}\, , 
\ee
where ${\bf 8}_v,{\bf 8}_{\pm}$ are the vector and the two spinorial 
representations of $SO(8)$.  
The $SO(8)$ generators $\Gamma^{ab}=\frac{1}{2}[\Gamma^a,\Gamma^b]$ 
satisfy the relations 
\be
\Gamma^{ab}=\psi^{abc}\Gamma^9\Gamma_{8c}\, ,~~~
\psi_{abcd}\Gamma^{ab}=-(4+2\Gamma^9)\Gamma_{cd}\, , 
 \label{8c}
\ee
where, as usual, $\Gamma^9=\left(\matrix{1&0\cr 0&-1}\right)$ is the chirality 
matrix. 
It is not then difficult to verify that the generators
\be
G^{\a\b}=\frac{3}{8}\left(\Gamma^{\a\b}+\frac{1}{6}{\Psi^{\a\b}}_{\g\d}
\Gamma^{\g\d}\right)\, , \label{G} 
\ee
leave the right-handed spinor $\eta_+$ invariant
\be 
G^{\a\b}\eta_+=0 
\, . \label{Gh}
\ee
The stability 
group of $\eta$ is again another $SO(7)$, which we will denote by 
$SO(7)^+$,
according to which  
\be
{\bf 8}_v={\bf 8}\, , ~~{\bf 8}_+={\bf 7}+{\bf 1}\, , 
~~{\bf 8}_-={\bf 8}
\, . \label{8+}
\ee
The singlet in the decomposition of ${\bf 8}_+$ corresponds to the null
eigenspinor $\eta$ of $G^{\a\b}$ in eq.(\ref{Gh}). This can be seen by 
considering the Casimir $G^{\a\b}G_{\a\b}$ which has a zero eigenvalue 
for the ${\bf 8}_+$. 
If we had used in eq.(\ref{G}) the second ${\bf 35}$ of $SO(8)$, the 
antiself-dual one, then $G^{\a\b}$ would annihilate the left-handed 
spinor $\eta_-$. In this case, the stability group of the latter 
is a  third $SO(7)^-$ subgroup of $SO(8)$ defined by
\be
{\bf 8}_v={\bf 8}\, ,  ~~{\bf 8}_+={\bf 8}\,,~~{\bf 8}_-={\bf 7}+{\bf 1}\, .
\ee
Again, $G^{\a\b}$ constructed with the antiself-dual ${\bf 35}$ has a 
zero eigenvalue which corresponds to the singlet in ${\bf 8}_-$. 
The three different $SO(7)$ subgroups of $SO(8)$, $SO(7)_v,
SO(7)^\pm$ are related by triality. 

\section{$Spin(7)$-Holonomy Spaces}

Let us now assume that the eight-dimensional spin manifold $X$
admits a spin connection which  satisfies the
self-duality condition
\be
 \omega_{\alpha\beta\mu}=\frac{1}{2}\Psi_{\a\b\g\d} 
{\omega^{\g\d}}_{\mu}
\, , \label{sd}
\ee
where $\Psi_{\a\b\g\d}$ is defined in eq.(\ref{p}).
It is not difficult to verify that a relation of the form
$\omega_{\alpha\beta\mu}= \lambda\Psi_{\a\b\g\d} 
{\omega^{\g\d}}_{\mu}$ is consistent for $\lambda=1/2,-1/6$ as can be 
verified by multiplying both sides 
by $\Psi^{\a\b\zeta\eta}$ and using eq.(\ref{fund}).
Here we consider only the case $\lambda=1/2$, while the
other possibility has not been investigated and left to
future work.
Eq.(\ref{sd}) is an eight-dimensional analog of the standard
self-duality condition in four dimensions  
where the totally antisymmetric symbol $\epsilon_{abcd}$ is replaced by
$\Psi_{\a\b\g\d}$.
Eight-dimensional
manifolds with a connection satisfying  
eq.(\ref{sd}) in the Euclidean regime 
will be called gravitational instantons. Note that the octonions have also 
appeared in the construction of an eight-dimensional Kerr metric \cite{CH}.  
The holonomy group
of such manifolds is in $Spin(7)$ as can be
seen as follows. Recall first that a manifold is of $Spin(7)$ holonomy 
if  and only if the 
Cayley four-form \cite{sal} 
\be 
\Psi&=&e^1\wedge e^2\wedge e^3\wedge e^8+ e^5\wedge e^1\wedge e^6\wedge e^8+
e^6\wedge e^2\wedge e^4\wedge e^8+e^4\wedge e^3\wedge e^5\wedge e^8\nonumber \\
&&+e^4\wedge e^7\wedge e^1\wedge e^8+e^6\wedge e^7\wedge e^3\wedge e^8
+e^5\wedge e^7\wedge e^2\wedge e^8+e^4\wedge e^5\wedge e^6\wedge e^7\nonumber\\
&&+e^2\wedge e^3\wedge e^7\wedge e^4+e^1\wedge e^3\wedge e^5\wedge e^7
+e^1\wedge e^3\wedge e^5\wedge e^7+e^1\wedge e^2\wedge e^7\wedge e^6 
\nonumber \\
&&+e^2\wedge e^3\wedge e^5\wedge e^6
+e^1\wedge e^2\wedge e^4\wedge e^5
+e^1\wedge e^3\wedge e^4\wedge e^6\, , \label{c}
\ee
where $e^\a,~ \a=1,\ldots,8$ is an orthonormal frame (the metric is 
$\sum_ie^i\otimes e^i$ in this frame), is torsion free, i.e., if it is closed 
\be
d\Psi=0\, . 
\ee
In this case, the manifold is Ricci flat.
The Cayley form  $\Psi$ is self-dual and, 
in addition, it is invariant under $Spin(7)$. It is the singlet in the 
decomposition of the ${\bf 35}_+$ in  the non-standard embedding  
of $SO(7)$  in $SO(8)$ given in 
eq.(\ref{8+})  according to which  
\be
{\bf 35}_v={\bf 35}\, , ~~~  {\bf 35}_+={\bf 1} +
{\bf 7}+{\bf 27}\, ,  ~~~ {\bf 35}_-={\bf 35}\, .
\ee

In order to prove now 
that manifolds which satisfy eq.(\ref{sd})
have $Spin(7)$ holonomy, let us observe that the Cayley four-form $\Psi$ 
can actually be written as 
\be
\Psi=\frac{1}{4!}\Psi_{\a\b\g\d}e^\a\wedge e^\b\wedge e^\g\wedge e^\d\, . 
\label{omega}
\ee 
It is not difficult to verify using eq.(\ref{fund}) and the structure 
equations 
$$de^\a+{\omega^\a}_\b e^\b=0\, , ~~~~~~{\omega^\a}_\b=
{\omega^\a}_{\b\mu} dx^\mu\,  $$
that if the spin connection satisfies 
eq.(\ref{sd}) the Cayley form $\Psi$ will be closed. Thus, manifolds 
with connection satisfying the self-duality condition (\ref{sd}) 
are $Spin(7)$ Ricci flat manifolds.
In addition,  such manifolds  admit
covariantly constant spinors. Indeed, for an $SO(8)$ spinor $\eta$ 
with
\be
D_\mu\eta=\left(\partial_\mu-\frac{1}{4}\omega_{\alpha\beta\mu}
\Gamma^{\alpha\beta}\right)\eta\, , \label{eta}
\ee
where $\alpha,\beta,...=1,...,8$ are world indices, $\mu,\nu,...=1,...,8$
are curved space indices and 
$\omega_{\alpha\beta}$ satisfy eq.(\ref{sd}),
one may easily verify that  $\eta$  is necessarily right-handed and 
\be
D_\mu\eta=
\partial_\mu\eta^A -\frac{1}{2}\omega_{\a\b\mu}G^{\a\b}\eta^A\, ,
\ee
where $G^{\a\b}$ has been defined in eq.(\ref{G}).  
Thus, $\eta$ is covariantly constant $D_\mu\eta=0$
if it is a constant spinor and satisfies eq.(\ref{Gh}).

Note also that the integrability condition of eq.(\ref{eta}) is
\be
R_{\a\b\mu\nu}\Gamma^{\a\b}\eta^A=0\, , 
\ee
from which follows, after multiplication with $\Gamma^\mu$, that
spaces of $Spin(7)$ holonomy are indeed Ricci flat.

\section{The Octonionic  Gravitational  Instanton}

We have just seen that a systematic way to construct manifolds 
of $Spin(7)$ holonomy is provided by solving
eq.(\ref{sd}). Selecting one direction, say the eighth,  
eq.(\ref{sd}) is written as 
\be
{\omega^8}_r=-\frac{1}{2}\psi_{rpq}\omega^{pq}\, . \label{seq}
\ee
The rest of the equations, namely,
\be
{\omega^p}_q=\frac{1}{2}{\psi^p}_{qrs}\omega^{rs}-
{\psi^p}_{qr}\omega^{r8}\, , \label{seq2} 
\ee
are automatically satisfied if eq.(\ref{seq}) holds. The self-duality 
conditions are then explicitly written as 
\be
\omega_{81}=-\left(\omega_{23}+\omega_{65}+\omega_{47}\right)\, , &&
\omega_{82}=-\left(\omega_{31}+\omega_{46}+\omega_{57}\right)\, ,\nonumber\\ 
\omega_{83}=-\left(\omega_{12}+\omega_{54}+\omega_{67}\right)\, , &&
\omega_{84}=-\left(\omega_{62}+\omega_{35}+\omega_{71}\right)\, ,\nonumber\\
\omega_{85}=-\left(\omega_{16}+\omega_{43}+\omega_{72}\right)\, , &&
\omega_{86}=-\left(\omega_{51}+\omega_{24}+\omega_{73}\right)\, ,\nonumber\\
\omega_{87}=-\left(\omega_{14}+\omega_{36}+\omega_{25}\right)\, ,  &&   
\label{expli}
\ee

We will assume that the metric of the $Spin(7)$-holonomy  manifold $X$
is of the form
\be
ds^2=f(r)^2dr^2+g(r)^2\left(d\mu^2+\frac{1}{4}\sin^2\mu
~\Sigma_i^2\right)
+h(r)^2\left(\sigma_i-A_i\right)^2\, , \label{an}
\ee
where $\Sigma_i$ and $\sigma_i,\, i=1,2,3$ are $SU(2)$ left-invariant 
one-forms satisfying
\be                                                       
d\Sigma_i=-\frac{1}{2}\epsilon_{ijk}\Sigma_j \wedge \Sigma_k,~~~~~
d\sigma_i=-\frac{1}{2}\epsilon_{ijk}\sigma_j \wedge \sigma_k,
\ee
and
\be
A_i=cos^2\frac{\mu}{2} ~ \Sigma_i\, .
\ee
In terms of the angular coordinates $\a,\b,\g,\theta,\phi,\psi$ with
$0<\a,\theta\leq\pi$, ~~$0<\b,\phi\leq 2\pi$, ~~$0<\g,\psi\leq 4\pi$, the
$SU(2)$ left-invariant one-forms $\Sigma_i,\sigma_i$ are
explicitly given by
\be
&\!\!\!&\Sigma_1=\cos\g~d\a+\sin\g\sin\a~d\b,
~~\Sigma_2=-\sin\g~d\a+\cos\g\sin\a~d\b ,
~~\Sigma_3=d\g+\cos\a~d\b , \nonumber \\
&\!\!\!&\sigma_1=\cos\psi~d\theta+\sin\psi\sin\theta~d\phi,
~~\sigma_2=-\sin\psi~d\theta+\cos\psi\sin\theta~d\phi ,
~~\sigma_3=d\psi+\cos\theta~d\phi , \nonumber
\ee

The surfaces $r=const.$ have the topology of $SU(2)$-bundle over
$S^4$ with $A_i$ the $k=1$ $SU(2)$ instanton on $S^4$.
The isometry of $X$ for generic $g(r),h(r)$
are $SO(5)\!\times\!SU(2)$. In particular, if $g(r)=h(r)$ the isometry
group is enlarged to $SO(8)$ since in this case the surfaces
$r=const.$ are round seven-spheres $S^7$ written in their Hopf-fibered
form. It should be noted that the ansantz for the metric (\ref{an})
is similar to the one for the Eguchi-Hanson 4d gravitational
instanton where the Hopf-fibration of the $S^7$ is replaced with the 
corresponding fibration of $S^3$.

Employing the orthonormal base
\be
e^i=\frac{1}{2}g(r)\sin\mu\,\Sigma_i\, , &&
e^{\hat{i}}=h(r)\left(\sigma_i-\cos^2\frac{\mu}{2}\,
\Sigma_i\right)\, , \nonumber\\
e^7=f(r)dr\, , && e^8=g(r)d\mu\, , \label{orthonormal}
\ee
where $i=1,2,3$ and $\hat{i}=4,5,6=\hat{1},\hat{2},\hat{3}$, we find
that the spin connection is
\be
\omega_{ij}=-\left(\frac{1}{g(r)\sin\mu}e^k+\frac{h(r)}{2g(r)^2}
e^{\hat{k}}\right)\epsilon_{ijk}\, , &&
\omega_{\hat{i}\hat{j}}=-\left(\frac{1}{2h(r)}e^{\hat{k}}+
\frac{2\cos^2\mu/2}{g(r)\sin\mu}e^k\right)\epsilon_{ijk}\, , \nonumber \\
\omega_{i\hat{j}}=\frac{h(r)}{2g(r)^2}\epsilon_{ijk}
e^k+\frac{h(r)}{2g(r)^2}\delta_{ij}e^8\, , && \omega_{8i}=
-\frac{1}{g(r)}\cot\mu\, e^i-\frac{h(r)}{2g(r)^2}e^{\hat{i}}\, ,
\nonumber \\
\omega_{8\hat{i}}=-\frac{h(r)}{2g(r)^2}e^i\, , &&
\omega_{87}=\frac{g(r)'}{g(r)f(r)}e^8\, , \nonumber \\
\omega_{i7}=\frac{g(r)'}{g(r)f(r)}e^i\, , &&
\omega_{\hat{i}7}=\frac{h(r)'}{h(r)f(r)}e^{\hat{i}} \, ,
\label{connection}
\ee
with $'=d/dr$. Then, using the components of the connection, we find that
 eq.(\ref{expli}) is satisfied if  $h(r),g(r)$ obey the 
differential equations 
\be
\frac{h(r)'}{h(r)f(r)}+\frac{1}{2h(r)}-\frac{h(r)}{g(r)^2}=0\, ,~~~~~
\frac{g(r)'}{g(r)f(r)}+\frac{3h(r)}{2g(r)^2}=0 \, . \label{equation}
\ee
We may use the reparametrization invariance of the metric under
$r\to r^\prime=r^\prime(r)$ to fix one of the three functions
$h(r),g(r)$ and $f(r)$. In particular, choosing
\be
g(r)^2=\frac{9}{20}r^2\, , \label{g}
\ee
we find after solving eq.(\ref{equation}) that
\be
h(r)^2=\frac{1}{5}g(r)^2\left(1-\left(\frac{m}{r}\right)^{10/3}\right)\,, ~~~~~~
f(r)^2=\frac{1}{1-\left(\frac{m}{r}\right)^{10/3}}\, , \label{gf}
\ee
where $m$ is an integration constant taken as the moduli. 
Thus, the full metric (\ref{an}) turns out to be
\be
ds^2\!=\! \frac{dr^2}{1\!-\!\left(\frac{m}{r}\right)^{10/3}}+
\frac{9}{20}r^2\left(d\mu^2
\!+\!\frac{1}{4}\sin^2\mu\, \Sigma_i\right)+\frac{9}{100}r^2
\left(1\!-\!\left(\frac{m}{r}\right)^{10/3}\right)\!\left(
\sigma_i-\cos^2\frac{\mu}{2}\,
\Sigma_i\right)^2\!. \label{metric}
\ee
It resembles the Eguchi-Hanson metric with moduli $m$, but in 
the present case we have a radial fall off rate $r^{-10/3}$, instead 
of $r^{-4}$ in Eguchi-Hanson,  
because of the eight-dimensional nature of our 
solution.   

One may easily verify that the apparent singularity at $r=m$ is a
removable {\it bolt} singularity. Indeed, near $r=m$, the metric
(\ref{metric}) at $\mu,\a,\b,\g=const.$, takes the form
\be
ds^2\sim
\frac{9}{25}\left(d\rho^2+\frac{1}{4}\rho^2\sigma_i^2\right)\, ,
\label{bolt}
\ee
with $\rho^2=r^2\left(1-(m/r)^{10/3}\right)$ which  is the
metric on ${\bf R}^4$ in polar coordinates. Thus, the topology of $X$
is locally of the form ${\bf R}^4\!\times\!S^4$ and the singularity at
$r=m$ is just a coordinate singularity.
Asymptotically  the metric (\ref{metric}) takes the form
\be
ds^2= dr^2+
\frac{9}{20}r^2\left(d\mu^2
+\frac{1}{4}\sin^2\mu\, \Sigma_i+\frac{1}{5}
\left(\sigma_i-\cos^2\frac{\mu}{2}\,
\Sigma_i\right)^2\right), \label{metric11}
\ee
and thus the boundary at infinity is the squashed $S^7$ \cite{DD}.

Finally we note for completeness that the bolt structure and the
asymptotic behaviour of the metric are insensitive to the value of
the free parameter $m$. This should be compared to the similar
behaviour of the Eguchi-Hanson metric in four dimensions, 
which for $m=0$ reduces 
to the flat space metric in polar coordinates modulo global issues
that arise from identifications in the allowed range of the Euler angles 
resulting in $S^3 / Z_2$ at infinity.    

\section{Brane Solutions}

The solution (\ref{metric}) we found above  
can be lifted to appropriate brane solutions in 
ten- and eleven-dimensional supergravity. In particular, the octonionic 
gravitational  instanton may be taken as the transverse space of a 
fundamental string in string theory or of a fundamental membrane in M-theory.
It can also be considered as a domain-wall solution (eight-brane) in massive 
IIA supergravity. In all cases, these brane solutions preserve 
the minimum amount of supersymmetry 
since there exists only one covariant constant spinor on the octonionic 
gravitational instanton. Thus, the brane solutions we will describe in 
the following are 1/16 BPS states.
\vspace{.5cm}

\noindent
$\bullet$ {\large{\it Type II Octonionic Fundamental String}}

\vspace{.3cm}

We consider ten-dimensional backgrounds of the form $M^2\times N^8$
with metric
\be
ds^2=e^{2A}(-dt^2+dx^2)+e^{2B}g_{mn}dy^mdy^n\, , 
\ee
where $y^m, m=1,...,8$ are coordinates on the transverse space $N^8$,
$A=A(y),~B=B(y)$ and $g_{mn}$ is the metric on $N^8$. 
Similarly, we assume that  the dilaton $\phi=\phi(y)$ and
the only non-vanishing component of the antisymmetric NS-NS two-form 
$B_{MN}$ is $B_{tx}=-e^{C(y)}$. Then, we find that the fermionic shifts 
equal to 
\be
\delta\psi_M&=&D_M\epsilon+\frac{1}{96}e^{-\phi/2}
\left({\Gamma_M}^{NPQ}-9{\delta_M}^N\Gamma^{PQ}\right)H_{NPQ}\,\epsilon
\nonumber \\
\delta\lambda&=&-\frac{1}{2\sqrt{2}}\Gamma^m\partial_m\phi\epsilon+
\frac{1}{24\sqrt{2}}e^{-\phi/2}\Gamma^{MNP}H_{MNP}\, \epsilon\, , \label{ssusy}
\ee
vanish if 
\be
A=\frac{3\phi}{4}\, , ~~~B=-\frac{\phi}{4}\, , ~~~C=2\phi\, ,
\ee
and the transverse space $N^8$ admits a covariant constant spinor. Finally, 
from the field equations, we 
obtain that $e^{-2\phi}$ is harmonic in $N^8$, i.e.,
\be
\nabla^2e^{-2\phi}=0\, .  \label{phi}
\ee

Recall at this point that our octonionic gravitational instanton 
admits a covariant constant spinor and so it can be 
taken as a solution for the transverse space $N^8$. 
In this case eq.(\ref{phi}) reduces to  
\be
\frac{1}{r^7}\partial_r\left(r^7\left[1-\left(\frac{m}{r}\right)^{10/3}
\right]
\partial_re^{-2\phi}\right)=0\, . 
\ee
The solution is described in terms of the Gauss 
hypergeometric function $_2F_1(a,b;c,z)$ as
\be
e^{-2\phi(r)}=e^{-2\phi_\infty}-\frac{\Lambda_2}{r^6}\,\,
{_2F_1}\left(1,\frac{9}{5};\frac{14}{5},(m/r)^{10/3}\right)\, , 
\ee
where $\Lambda_2$ is a constant proportional to the string tension $T_2$, 
using the integral representation of  $_2F_1(a,b;c,z)$
\be
_2F_1(a,b;c,z)= \frac{\Gamma(c)}{\Gamma(b)\Gamma(c-b)}\int_0^1
t^{b-1}(1-t)^{c-b-1}(1-zt)^{-a}dt\, , ~~~~~{\it Re}\,c>{\it Re}\,b>0\, .
\ee

To compare it with other results in the literature, we note 
that a heterotic octonionic string was constructed in 
\cite{HS} with (0,1) supersymmetry on its world-volume. In that case, 
however, the Bianchi identity for   
$H_{MNP}$ was solved by introducing appropriate self-dual 
eight-dimensional gauge fields; for the relevant technical 
details we refer the reader to the literature \cite{FNG}. 
Hence, the solution we describe in this paper is different.  
 
\vspace{.5cm}

\noindent
$\bullet$ {\large {\it Octonionic Fundamental Membrane}}

\vspace{.3cm}

We may also lift the octonionic gravitational instanton in 
the eleven-dimensional supergravity. Here, the bosonic fields in the 
graviton multiplet is  the 
graviton and the antisymmetric three-form $A_{MNP}$. We will thus 
consider an eleven-dimensional background of the form $M^3\times 
N^8$ with metric 
\be
ds^2=e^{2A}(-dt^2+dx_1^2+dx_2^2)+e^{2B}g_{mn}dy^mdy^n\, , 
\ee
and non-vanishing components for $A_{MNP}$, 
namely $A_{tx_1x_2}=e^{-C}$, where 
$A=A(y),~B=B(y)$ and  $C=C(y)$ as before.  
In this case, the gravitino shift 
\be
\delta\psi_M&=&D_M\epsilon+\frac{1}{288}
\left({\Gamma_M}^{NPQR}-8{\delta_M}^N\Gamma^{PQR}\right)F_{NPQR}\,\epsilon
\ee
vanish if there exists a covariant constant spinor on $N^8$ and 
\be
A=\frac{2}{3}C\, , ~~~B=-\frac{1}{3}C\, .
\ee
Thus, we may choose for $N^8$ the octonionic 
gravitational instanton as before. In addition,
the field equations are satisfied if $e^{-C}$ is a harmonic function on $N^8$.
In this case we find 
\be
e^{-C(r)}=
e^{-C_\infty}-\frac{\Lambda_3}{r^6}\,\,
{_2F_1}\left(1,\frac{9}{5};\frac{14}{5},(m/r)^{10/3}\right)\, , 
\ee
where now $\Lambda_3$ is proportional to the membrane tension $T_3$. 

The fundamental octonionic membrane we have just 
constructed is completely different from 
the octonionic membrane constructed in \cite{DL}.
We only note that the latter is not supersymmetric and 
furthermore it has a non-trivial $F_{mnpq}$
proportional to the self-dual tensor in eq.(\ref{p}). 
Hence our result yields a new configuration in 
eleven-dimensional supergravity. 

\section{Discussion}

Summarizing, we have shown that eight-dimensional gravitational 
instantons can be constructed 
as solutions of self-duality equations for the spin 
connection. The octonions, which are integral part of our 
method, yield in this fashion Ricci flat manifolds with holonomy in 
$Spin(7)$. We have also presented an explicit solution with 
isometry group $SO(5) \times SU(2)$ having a bolt singularity and
the asymptotic structure of the squashed seven-sphere; it can be
regarded as a higher dimensional analogue of the Eguchi-Hanson 
space. Also, because of supersymmetry, we can elevate our solution 
to ten and eleven dimensions in order to obtain octonionic 
fundamental string and membrane configurations respectively 
which are 1/16 BPS states.

The present framework can be developed further by considering 
eight-dimensional solutions with $SU(2)$ isometry and attempt
a classification of the corresponding gravitational instanton 
symmetries in analogy with four dimensions \cite{GHP}. In the
four dimensional case one considers the action of an 
one-parameter isometry group $U(1)$ and classifies its fixed 
points into two different types: isolated points called nuts
and 2-surfaces (spheres) called bolts. Then, performing a dimensional
reduction of Einstein's equations with respect to the orbits
of the $U(1)$ isometry group to three dimensions, 
one exhibits a duality between the
electric and the magnetic aspects of gravity (corresponding 
to bolts and nuts respectively) via the Ehlers transform.
Of course, certain solutions might exhibit
more isometries which are not of general value; for 
instance the Eguchi-Hanson space has an additional $SO(3)$
isometry apart from the $U(1)$. 
In eight dimensions, although the particular solution 
we found has 
an additional $SO(5)$ isometry apart from $SU(2)$, we think 
that focusing on more general classes of solutions with $SU(2)$ 
isometry are worth studying further. Selecting gravitational
instantons with $SU(2)$ isometry in eight dimensions is 
also quite natural from the point of view of Hopf fibrations,
as is the selection of gravitational instantons with $U(1)$   
isometry in four dimensions.

In the reduction of an eight dimensional theory to five 
dimensions using the orbits of the corresponding isometry
group, $S^3$, we are naturally led to the notion of 
``nut potential" as an antisymmetric 2-form field that dualizes
the $SU(2)$ gauge field connection of the metric. The reasoning is
the same one that yields an axion-like scalar function as nut
potential in four-dimensional spaces with $U(1)$ isometry.
It will be interesting to carry out this line of investigation
in the present case and see whether $M2$ and $M5$ branes, when
appropriately taken in eleven-dimensional supergravity, provide
the analogue of bolts and nuts in establishing various 
electric and magnetic aspects of eight dimensional gravitational
instantons. We hope that in this way, apart for improving our
conceptual understanding of the general problem, we will also
be able to find more explicit instanton solutions as in four dimensions.
We plan to report on these issues elsewhere. 
\vspace{.5cm}

\noindent
{\bf Note Added:} We have been informed by B. Acharya (private communication) 
that he has also employed the squashed $S^7$ in searching for 
$Spin(7)$-holonomy manifolds (unpublished work). We also became aware 
after circulating the first draft of our paper that the particular metric 
(37) was constructed some time ago by solving directly Einstein's equations 
in eight dimensions \cite{GPP}; we thank G. Gibbons and C. Pope for bringing 
their paper to our attention. Their study was based on second order 
differential equations rather than the first order self-duality equations (18),
and (in a sense) made explicit    earlier mathematical results on the 
construction of complete metrics with $Spin(7)$ holonomy \cite{BS}. We thank
 D. Morrison
for also pointing out the relevance of ref.\cite{BS} to the general framework 
we have adopted here.

\end{document}